\documentclass[conference]{IEEEtran}
\input epsf
\usepackage{graphicx}
\usepackage{tabularx}
\usepackage[table,x11names]{xcolor}
\usepackage{array}
\usepackage{hhline,colortbl}
\usepackage{array}
\usepackage{booktabs}
\usepackage{subcaption}
\usepackage{multirow}
\usepackage{url}
\definecolor{LightCyan}{rgb}{0.88,1,1}
\definecolor{Gray}{gray}{0.5}
\newcommand{\botometer}{$\mathsf{BotoM}$}
\newcommand{\spambot}{$\mathsf{SpamBot}$}
\newcommand{\ndsync}{$\mathsf{NDSync}$}
\newcommand{\score}{$\mathsf{SCoRe}$}
\newcommand{\fakeacc}{$\mathsf{FakeAcc}$}
\IEEEoverridecommandlockouts

\usepackage{amsmath}   
\RequirePackage[textsize=scriptsize]{todonotes}
\begin{document}

\title{\LARGE Retweet Us, We Will Retweet You:\\ Spotting Collusive Retweeters Involved in Blackmarket Services
\thanks{{\color{blue}This paper has got accepted in IEEE/ACM International Conference on Advances in Social Networks Analysis and Mining
(ASONAM 2018)}}
}

\author{\IEEEauthorblockN{Hridoy Sankar Dutta}
\IEEEauthorblockA{IIIT Delhi, India\\
hridoyd@iiitd.ac.in}
\and
\IEEEauthorblockN{Aditya Chetan$^*$}\thanks{$^*$Equal contribution}
\IEEEauthorblockA{IIIT Delhi, India\\
aditya16217@iiitd.ac.in}
\and
\IEEEauthorblockN{Brihi Joshi$^*$}
\IEEEauthorblockA{IIIT Delhi, India\\
brihi16142@iiitd.ac.in}
\and
\IEEEauthorblockN{Tanmoy Chakraborty}
\IEEEauthorblockA{IIIT Delhi, India\\
tanmoy@iiitd.ac.in}
}

%


\maketitle

\begin{abstract}
Twitter has increasingly become a popular platform to share news and user opinion. A tweet is considered to be important if it receives high number of affirmative reactions from other Twitter users via \verb|Retweets|. {\em Retweet count} is thus considered as a surrogate measure for positive crowd-sourced reactions -- high number of retweets of a tweet not only help the tweet being broadcasted, but also aid in making its topic trending. This in turn bolsters the social reputation of the author of the tweet. Since social reputation/impact of users/tweets influences  many decisions (such as promoting brands, advertisement etc.), several blackmarket syndicates have actively been engaged in producing fake retweets in a collusive manner. Users who want to boost the impact of their tweets approach the blackmarket services, and gain retweets for their own tweets by 
retweeting other customers' tweets. Thus they become customers of blackmarket syndicates and engage in fake activities. Interestingly, these customers are neither bots, nor even fake users -- they are usually normal human beings; they {\em express a mix of organic and inorganic retweeting activities}, and {\em there is no synchronicity across their behaviors}.

In this paper, we make a first attempt to investigate such blackmarket customers engaged in producing fake retweets. We collected and annotated a novel dataset comprising of customers of many blackmarket services and characterize them using a set of 64 novel features. We show how their social behavior differs from genuine users. We then use state-of-the-art supervised models to detect three types of customers (bots, promotional, normal) and genuine users. We achieve a Macro F1-score of $0.87$ with SVM, outperforming four other baselines significantly. We further design a browser extension, \score~ which, given the link of a tweet, spots its fake retweeters in real-time. We also collected users' feedback on the performance of \score~ and obtained 85\% accuracy.   
 
\end{abstract}
\IEEEoverridecommandlockouts

%
\IEEEpeerreviewmaketitle

\section{Introduction}
Twitter, arguably the most popular micro-blogging site, provides its users two major ways to place  their affirmative reactions towards different entities: (i) user-level affirmation (such as \verb|follow|), and (ii) content-level affirmation (such as \verb|retweet|, \verb|like|). Here, we particularly focus on `retweeting activity', a major content-level affirmative action, which provides a way of re-broadcasting messages and confirms retweeter's agreement of the  message being  important to others. Retweet count of a tweet gives the OSN users a sense of crowdsourced agreement on the tweet, and thus determines the influence of the tweet as well as the author of the tweet. It also helps in making a certain topic trending on Twitter. This further brings an opportunity of `intentional manipulation' to the social adversaries who falsely create the impression of popularity of tweets by generating huge volume of retweets.

\begin{figure}[!t]
\centering
\begin{subfigure}[b]{0.24\textwidth}
\includegraphics[width=\textwidth]{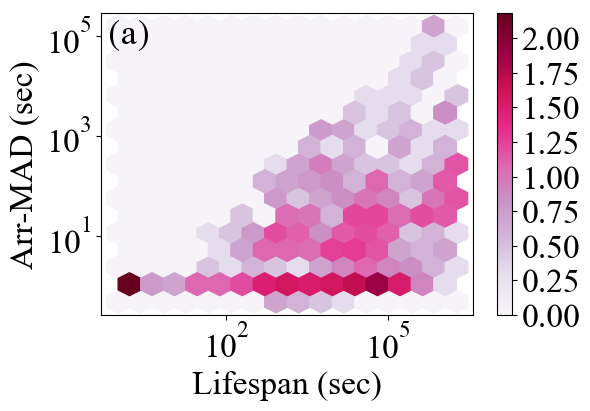}
\label{fig:followerCDF}
\end{subfigure}
\begin{subfigure}[b]{0.24\textwidth}
\includegraphics[width=\textwidth]{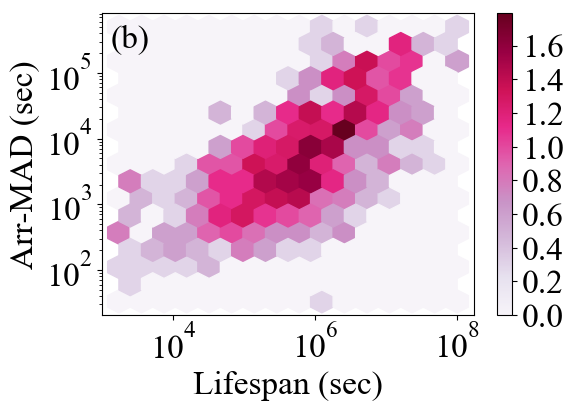}
\label{fig:fbyf1}
\end{subfigure}

\caption{{\bf (b) Asynchronous behavior of collusive retweeters  as opposed to the (a) synchronicity of normal retweet fraudsters mentioned in \cite{Neil:ndsync}}. The figure shows two binned 2-D heatmaps (in logarithmic scale) of {\em Lifespan} (time elapsed between the first and last retweets)  vs. {\em Arr-MAD} (mean absolute deviation of retweets’ inter-arrival times) of retweet threads for (a) normal retweet fraudsters (data taken from \cite{Neil:ndsync}) and (b) collusive retweeters. Unlike (a) where {\em Arr-MAD} is almost invariant with {\em Lifespan}, (b) shows an increasing trend.  }\label{fig:motivation}
\end{figure}

\begin{table}
    \centering
    \caption{{\bf Macro F1-score of the competing methods.} None of the existing methods can detect  collusive retweeters accurately. }\label{tab:summery}
    \begin{tabular}{|c|c|c|c|}
    \hline
    {\bf Bot}    & {\bf Fake Account} & {\bf Sync. Fake Retweeter}  & {\bf Our}  \\
    {\bf Detection \cite{davis2016botornot}} & {\bf Detection \cite{elazab2016fake}} & {\bf Detection \cite{Neil:ndsync}} & {\bf Method} \\\hline
     0.760 & 0.693 & 0.750 & 0.873 \\\hline
    \end{tabular}
    \vspace{-5mm}
\end{table}

{\bf Challenges in detecting collusive retweeters:} There exist several blackmarket agencies which have created thriving and intelligent ecosystems of producing spam retweets. Users can gain retweet count
of their own tweets 
for free by retweeting tweets of other customers of those services. Thus users unwillingly become customers of blackmarket services. Such trend of inflating social reputation has become prevalent across different social media platforms. Detection of such blackmarket retweeters is challenging for many reasons -- (i) They are not bots, but human beings; therefore, bot detection algorithms can not flag them (first column of Table \ref{tab:summery}). (ii) Their Twitter accounts are not fake; therefore, fake account detection algorithms can not detect them (second column of Table \ref{tab:summery}). (iii) They express a {\em mix of organic and inorganic behavior} in their retweeting patterns -- they organically reweet some genuine tweets; at the same time, they inorganically retweet tweets submitted to blackmarket services. The extent of inorganic activities may differ across retweeters. (iv) They do not show any synchronicity across their retweeting patterns, thus making it difficult for the existing synchronous fake retweeter detection  methods \cite{Neil:ndsync} to detect them (third column of Table \ref{tab:summery}). 

{\bf Motivation and state-of-the-art:} Despite previous efforts in understanding different fraudulent activities in Twitter such as fake account detection \cite{elazab2016fake}, social spam detection \cite{wang2010don}, 
content affirmation via retweeting has hardly been studied. A preliminarily attempt was made by \cite{Neil:ndsync} to detect spam retweeters with the hypothesis that suspicious activities are highly `synchronized', i.e., a group of spammers retweet at the same time. However, we clearly observe that {\bf synchronicity does not hold among collusive retweeters} (Figure \ref{fig:motivation}) -- this is intuitive since the blackmarket services do not have control on the customers involved in producing fraud retweets. 
A very recent effort was made by \cite{anupama} to detect collusive followers in Twitter. They concluded that crowdsourced manipulation is a big threat to the
credibility of OSNs. {\bf However, to the best of our knowledge, ours is the first attempt to detect blackmarket-based collusive manipulation of content credibility via fraudulent retweeting activities.} Interestingly, Twitter has not yet been successful in flagging these customers -- we observed 31 blackmarket customers who are marked as `verified users' by Twitter.

{\bf Our Contributions:} We begin our experiment by surveying different types of blackmarket services (Section \ref{sec:blackmarket-services}) and collecting customers from multiple blackmarket services
(Section \ref{sec:data-collection}). 
We employed human annotators to label each customer into one of the following categories -- \textit{bots}, \textit{promotional} and \textit{normal}. We also collected genuine users who are researchers / experts in machine learning, following the method in \cite{shah2017many}.  The timeline information of these users (customers and genuine users) were further scraped. These two sets of users thus form the retweeter set which we further analyze. 
We then propose an exhaustive set of  64 novel features to characterize retweeters
(Section \ref{sec:experimental-setup}). We run several state-of-the-art supervised models to classify retweeters into four types -- bot, promotional customer, normal customer, genuine user. We observe that Logistic Regression achieves the maximum accuracy, outperforming the baseline methods significantly (Section \ref{sec:res_multi}). Our method (with SVM) turns out to be equally successful in categorizing retweeters into binary classes -- customers and genuine (Macro F1-socre of $0.87$). 

We finally build a browser extension, \score, `{\bf S}potting {\bf Co}llusive {\bf Re}tweers', which, given a retweeter set of a tweet, classifies  each retweeter as customer or genuine. The extension allows the end users to judge the quality of the classification \score~ produces and provide correct labeling, in case it produces wrong result. The user-generated feedback is collected in the back-end and the supervised model is retrained in an incremental way. However, special care has been taken to ignore spurious feedbacks (more details in Section \ref{sec:system}).
We seek feedback of 25 volunteers on the performance of \score~ and obtain 85\% accuracy. We believe that the widespread use of such extension would help OSN administrators block the customers and measure the true credibility of a tweet based on its genuine retweet count. 

\textbf{Reproducibility:} All the codes and processed dataset are available at {\color{blue}\url{https://tinyurl.com/y7ruh2o8}}.

\section{Related Work}


\textbf{Detection of fraudsters in OSNs:} 
Many studies have been conducted on detecting frauds in Twitter and other OSNs. \cite{benevenuto2010detecting} detected spammers in Twitter using features generated from tweet content and user social behavior. 
\cite{lee2015will} identified strangers on Twitter, who can be potential retweeters to help effectively propagate intended information within a desired time frame. \cite{s} focused on detecting inorganic retweet behavior and proposed `RTGEN', a realistic generator that imitates the behaviors of both honest and fraudulent users. \cite{chu2012detecting} used machine learning approach to classify Twitter accounts into  `human', `bot' and `cyborg'. \cite{ferrara2016rise} discussed the rise of Social Bots on Twitter. 
\cite{hu2013social} used network properties to identify spammers on Twitter. \cite{wang2010don} proposed a spam detection technique to identify suspicious users on Twitter. 
\cite{thomas2011suspended} studied multiple spamming techniques, including creating fake Twitter accounts, generating spam URLs, and spam distribution. 
\cite{jiang2014detecting} used synchronous and abnormal behaviors of Twitter followers to detect suspicious following behavior.
\cite{GuptaKGK018} detected spam campaigns on Twitter.

{\textbf{Detection of blackmarket customers in Twitter:}} Understanding the dynamics of blackmarket services have recently gained substantial attention among researchers. 
\cite{de2013twitter} provided a detailed analysis
of blackmarket services and their impact on multiple OSNs. Most of the prior works showed how fake followers in social media help in promoting a certain user \cite{stringhini2013follow}. 
\cite{shah2017many} studied multiple types of blackmarket agencies by showing their multifaceted behavior. \cite{thomas2013trafficking} studied various blackmarkets tied to fraudulent Twitter credentials, monitoring pricing, availability, and fraud perpetrated by these services.  
\cite{liu2016pay} tackled the problem of voluntary following activity detection  to automatically detect malicious accounts that make profit in the follower markets. \cite{motoyama2011analysis} performed an analysis of six different underground forums and showed how OSN users are engaged in selling goods and services.  

\textbf{Remarks.} Our methodology differs from the existing approaches in many ways: (i) \cite{Neil:ndsync} identified fake retweeting behavior in Twitter on the basis of synchronicity. This type of behavior can only be observed if the retweeters are controlled by a centralized authority. However, collusive blackmarket customers are not controlled by anyone. They intentionally retweet other customers' tweets. Therefore, they do not exhibit any synchronized behavior. 
(ii) Existing fraud detection approaches classify a user into `fraud' or `genuine' \cite{shah2017many,Neil:ndsync}. Our work classifies users into four categories -- `genuine', `bots', `promotional customers' and `normal customers'.
(iii) We are also the first to develop a real-time system that can identify customers and genuine retweeters, given the retweeter list of a tweet. 

\section{Blackmarket Services}
In this section, we describe our efforts in collecting and annotating a set of Twitter accounts corresponding to the customers of various blackmarket agencies. 

\subsection{Types of Blackmarket Retweet Services}  \label{sec:blackmarket-services} While investigating the mode of blackmarket services, we noticed that there are two prevalent models of services \cite{shah2017many} -- (i) \textbf{\textit{ Premium Services}} which only provide services upon receiving payment from the customers; (ii) \textbf{\textit{Freemium Services}} which, similar to premium services, offer both paid services, as well as unpaid services that require the users to provide their Twitter login details; this in turn may involve the users unconsciously in the blackmarket activities. Here our primary focus is to understand the {\em activities of freemium services} which are easy to access due to their unpaid service model. This type of services can further be divided into three categories:
\begin{itemize}
\item{\textbf{Social-share services:}} These services (e.g., FreeFollowers\footnote{\label{note1}FreeFollowers: \url{https://www.freefollowers.io/}, Like4Like: \url{https://like4like.org/}, TraffUp: \url{http://traffup.net/}, JustRetweet: \url{http://justretweet.com}.}) ask customers to perform activities on social-media contents of other customers involved in those services. The activities on social-media contents can be `Facebook Share', `Facebook Like', `Twitter Retweets' etc. 
\item{\textbf{Credit-based services}}: These services (e.g., Like4Like, YouLikeHits, TraffUp, JustRetweet) work on credit-based policies. Each customer has to put a value for the tweet, which gets deducted from his/her credits when the tweet is published. A customer retweets the tweets of other customers to earn credits and the value of each tweet is added up to his/her total credit.
\item{\textbf{Auto-time retweet services}}: The customers of these services (e.g., TweetsTool\footnote{http://www.tweetstool.com/}) need to get access token from Twitter and login to the service. They can request for 10-50 retweets for each of their tweets in a 15-minute window. Other than the retweet service, it also provides the customers with the following services -- `Auto Favourite', `Auto Follower', `Auto Reply'.
\end{itemize}

\subsection{Data Collection}  \label{sec:data-collection}
{\bf Collecting blackmarket customers:} In order to collect the  information of  blackmarket customers, we focused our crawling only on credit-based services because -- (i) their service policies are easy to understand, (ii) as opposed to the social-share services, they only utilize a single platform (Twitter) to perform their activities, (iii) most of the freemium services follow credit-based strategy compared to the other strategies which would help us in collecting more data for our analysis. We adopted an `active probing' strategy to collect the dataset -- we created multiple dummy accounts in each of the credit-based freemium services (Like4Like, YouLikeHits, TraffUp, JustRetweet), kept retweeting tweets of other customers whose tweets were posted on the blackmarket services, and collected their IDs. We continued these activities for one month (Feb, 2018). 
Table \ref{table:data_statistics} shows the statistics of the dataset for each service.  During this process, Twitter suspended some of the accounts we collected, and therefore we ignored those accounts for further analysis.
From the collected dataset, we noticed that the average number of tweets posted by a user to YouLikeHits is $1.43$ which is relatively high  compared to Like4Like ($1.01$). 
 Interestingly, we found a significant number of customer overlap (76 customers) across YouLikeHits and Like4Like. It suggests that customers utilize several such blackmarket services audaciously without any obfuscation of their identities to promote their tweets. We further scraped the timeline information of these customers using Twitter's REST API.

{\textbf{Collecting genuine users:}} To compare  normal users with blackmarket customers in Twitter, we selected $1000$ genuine users (following  \cite{shah2017many}) -- (i) We searched for Twitter lists containing users who are researchers, working in the field of Machine Learning/Data Mining/Information Retrieval. 
(ii) We also discarded genuine users who have more than 1 million followers since high follower count may resemble a celebrity and we wanted to discard any celebrity-like users from our analysis in order to remove any unnecessary bias. We further scraped the timelines of the remaining users.


\begin{table}[!t]
\caption{Statistics of the dataset.}
\label{table:data_statistics}
\vspace{-5mm}
\begin{center}
\begin{tabular}{{| >{\centering\arraybackslash}m{0.5in} | >{\centering\arraybackslash}m{0.3in} | >{\centering\arraybackslash}m{0.5in} | >{\centering\arraybackslash}m{0.5in} |
>{\centering\arraybackslash}m{0.6in} | 
}}
\hline
\centering \textbf{Service} & \textbf{\# users} & \textbf{\# users suspended / deleted} & \textbf{\# users taken for our analysis} & \textbf{\# tweets our dummy accounts retweeted}\\
\hline  YouLikeHits & 638 & 168 & 470 & 914 \\
 Like4Like & 451 & 178 & 273 & 459 \\
 TraffUp & 2 & 0 & 2 & 6 \\ 
JustRetweet & 11 & 3 & 8 & 1\\ 
\hline Genuine Users & 1000 & 0 & 1000 & -\\
\hline
\end{tabular}
\end{center}
\vspace{-5mm}
\end{table}






\subsection{Human Annotation}
Three human annotators\footnote{They were experts in social media, and their age ranged between 25-35.} were asked to label the blackmarket customers into  \textit{Bots}, \textit{Promotional Customers} and \textit{Normal Customers} based on our definition of each customer type and Twitter's terms of service. Annotators were also given freedom to search for any information related to customers and apply their own intuition. Here we describe the information we provided to the annotators about three types of customers: 

\begin{enumerate}
\item{\textbf{Bots:}} A Twitter Bot is a software which controls a Twitter account using the Twitter API \cite{chu2012detecting}. Bots on Twitter perform both helpful and harmful activities working in a coordinated fashion \cite{davis2016botornot}. 
\item{\textbf{Promotional customers:}} Twitter has always been a source of advertising content to its target users. Promotional users are involved in promoting one of the three products: \textit{tweets}, \textit{accounts} or \textit{services} \cite{eftekhar2013some}. The annotators were asked to search for tweets in the timeline of customers promoting one of the above products. We observed that many customers in our dataset were involved in promoting brands using keywords such as `win', `ad', `Giveaway'. 

\item{\textbf{Normal customers:}} These customer do not fall under any of the two categories mentioned above. 

\end{enumerate}

Each annotator was given all the customer accounts for annotation. Finally, we considered only $743$ customers whose labels were agreed upon by at least two annotators, and found 86 bots, 275 promotional customers and 382 normal customers. The average inter-annotator agreement was $0.79$ based on  Cohen's $\kappa$. 

\subsection{Interesting Observations}
\begin{itemize}
\item Customers of the blackmarket services often remove their tweets after a small duration. We found that around 3\% customers have removed their tweets.
\item We also found 1\% customers who are suspended by Twitter. This clearly indicates that Twitter is still inefficient in identifying customers of the blackmarket services.
\item We found 31 customers who are marked as verified genuine accounts by Twitter. This also indicates the inability of current Twitter policy to flag blackmarket customers.  
\item The Twitter profiles of 4\% customers are shown with the warning \textit{``Caution: This account is temporarily restricted"}. Twitter enforces this warning on users in regards to either of the following cases \footnotetext{https://maximizesocialbusiness.com/why-your-twitter-account-may-be-restricted-1169/}: (i) repeatedly posting duplicate or near-duplicate content, (ii) abusing trending topics or hashtags, (iii) sending automated tweets or replies, and (iv) using bots or applications to post similar messages based on keywords.
, (v) posting similar messages over multiple accounts, and  (vi) aggressively following and un-following people.
 \end{itemize}

\section{Experimental Setup} \label{sec:experimental-setup}
In this section, we present novel features to characterize different types of users (both customers and genuine users) and supervised models for classifying users.  

\subsection{Feature Selection}
We use an extensive set of 64 novel features to detect different types of users. We group these features into five buckets: (i) Profile Features (PF), (ii) Social Network Features (SNF), (iii) User Activity Features (UAF), (iv) Likelihood Features (LF), and (v) Fluctuation Features (FF).\\

\begin{figure}[!t]
      \centering
      \includegraphics[scale=0.42]{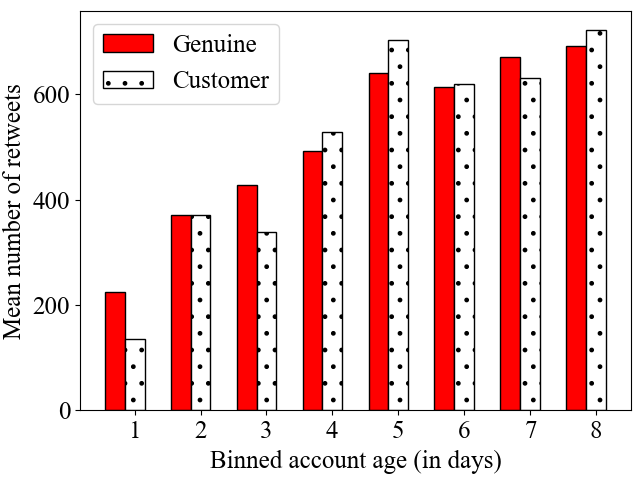}
      \caption{Relationship between retweet count and account age (in days; further divided into bins: bin 1: 0-499, bin 2: 500-999,$\cdots$, bin  8: 3500-3999).}
      \label{fig:accountAge}
      \vspace{-5mm}
\end{figure}

\noindent\textbf{(i) \underline{Profile Features (PF):}} We hypothesize that the older the account of a user, s/he will have more tendency to retweet others' tweets. Therefore, we take \textit{account age} ({\bf PF\textsubscript{1}}) as one feature  in our study. Figure \ref{fig:accountAge} corroborates our hypothesis by showing that both genuine users and customers exhibit an increasing trend of retweet count w.r.t their account age.
We further consider four other profile features: ({\bf PF\textsubscript{2}}) \textit{length of the screen name}, ({\bf PF\textsubscript{3}}) \textit{whether the profile has a description or not}, ({\bf PF\textsubscript{4}}) \textit{length of profile description}, ({\bf PF\textsubscript{5}}) \textit{whether profile has a URL or not}. \\

\noindent{\textbf{(ii) \underline{Social Network Features (SNF):}}} Social network features such as \textit{number of followees} ({\bf SNF\textsubscript{1}}) {\em and followers} ({\bf SNF\textsubscript{2}}) and {\em their ratio} ({\bf SNF\textsubscript{3}}) sometimes provide indicative information about the users. Figure \ref{fig:socialfeature}(a) shows that customers tend to follow a lot of other users, compared to the genuine users -- this may be explained by the typical tendency of customers to follow others with the intention that the followees will further draw  attention to their profiles as well as their tweets. Figure \ref{fig:socialfeature}(a:inset) shows that followee count has a more positive correlation (Spearman's $\rho=0.58$) with retweet count in case of genuine user than the customers (Spearman's $\rho=0.42$). 
Figure \ref{fig:socialfeature}(b) shows that genuine users tend to have a variety of follower count, whereas all customers have a smaller follower count. Figure \ref{fig:socialfeature} (b:inset) shows that genuine users with a large number of followers tend to retweet more compared to those with less followers; however, it may not hold for the customers. 
The ratio of followees to followers in Figure \ref{fig:socialfeature}(c) shows that genuine users have a lower followee to follower ratio -- this is expected as they tend to have around equal number of followees and followers (recall that we excluded celebrities from our study, whose follower count may be much higher than the followee count). For genuine users, we also observe a declining trend of retweet count with the increase of followee to follower ratio; however customers do not follow any such trend (Figure \ref{fig:socialfeature} (c:inset)).  
 We further measure the influence score of users using {\em Klout score}\footnote{http://klout.com} ({\bf SNF\textsubscript{4}}) and take it as a feature. Klout score returns a value between 1-100 to rate a user based on his/her online social influence. The Klout score of genuine users and customers based on their influence is shown in Figure \ref{fig:socialfeature}(d). The median value   of Klout score   for customers and genuine users is  48.11 and 43.57 respectively -- such small difference in Klout score indicates that though customers use blackmarket services, it does not help them making their profile significantly popular. Fig. \ref{fig:socialfeature} (d:inset) shows that the retweet count tends to increase with the increase of Klout score for genuine users; however there is not such uniform pattern for customers. \\

\noindent{\textbf{(iii) {\underline{User Activity Features (UAF):}}} User activity plays a significant role - higher the activity of a user, higher is the chance of his/her tweet to be retweeted by a stranger. We use the following features to measure the activity of users: ({\bf UAF\textsubscript{1}}) \textit{total number of tweets}, ({\bf UAF\textsubscript{2}}) {\em number of direct mentions per tweet}, ({\bf UAF\textsubscript{3}}) {\em number of URLs per tweet}, ({\bf UAF\textsubscript{4}}) {\em number of hashtags per tweet}, ({\bf UAF\textsubscript{5}}) {\em number of tweets per day}, ({\bf UAF\textsubscript{6}}) {\em number of retweets per day}, and ({\bf UAF\textsubscript{7}}) {\em number of retweets per tweet}. To compute these features, we crawl all (max 3200) tweets from the timeline of each user. We also use \textit{Bot-score} ({\bf UAF\textsubscript{8}}) of each user from Botometer service\footnote{\url{https://botometer.iuni.iu.edu}} \cite{davis2016botornot} and consider it as a feature.\\

\begin{figure}[t] 
  \begin{minipage}[b]{0.5\linewidth}
    \centering
    \includegraphics[width=.99\linewidth]{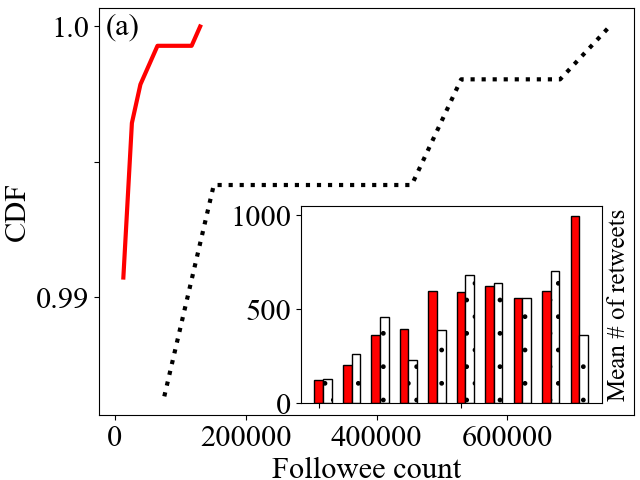} 
    \label{followeeCDF}
  \end{minipage}
  \begin{minipage}[b]{0.5\linewidth}
    \centering
    \includegraphics[width=.99\linewidth]{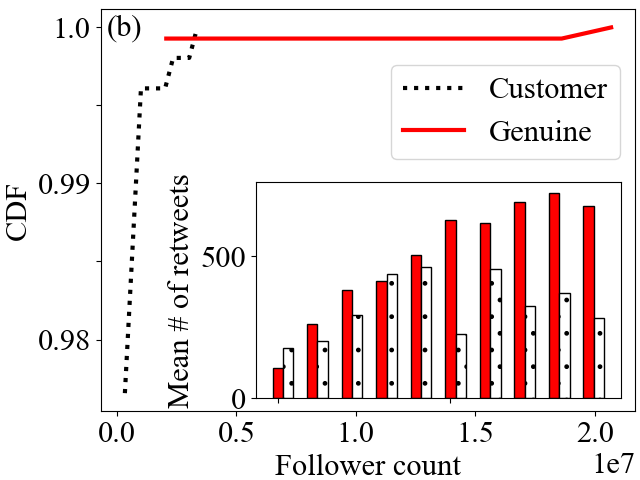}
    \label{followerCDF}
  \end{minipage} 
  \begin{minipage}[b]{0.5\linewidth}
    \centering
    \includegraphics[width=.99\linewidth]{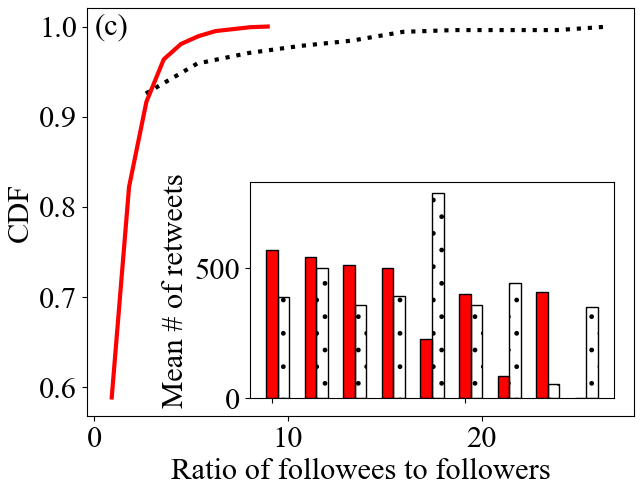} 
    \label{fig:fbyf}
  \end{minipage}
  \begin{minipage}[b]{0.5\linewidth}
    \centering
    \includegraphics[width=.99\linewidth]{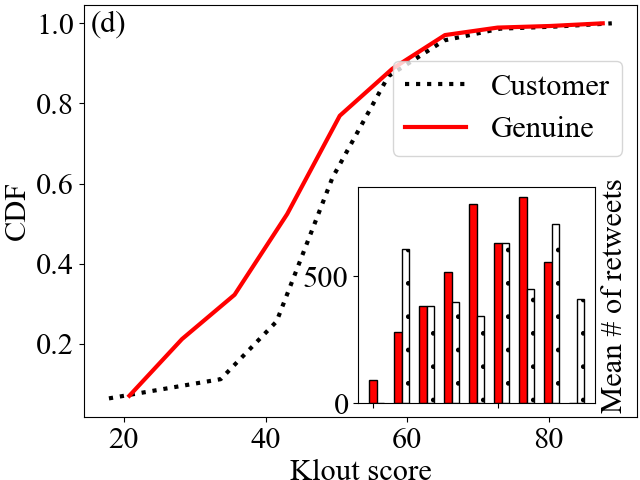} 
    \label{fig:klout}
  \end{minipage} 
  \caption{CDF of social network features -- (a) followee count, (b) follower count, (c) followee to follower ratio, (d) Klout score, and their relation with the retweet count (inset). The variable in x-axis is same for the main diagram and its inset.}\label{fig:socialfeature}
  \vspace{-7mm}
\end{figure}

\begin{table*}[!t]
\centering
\caption{Performance of different competing methods for multi-class classification.}\label{table:multiclassresult}
\begin{tabular}{|l||cccc||cccc|}
\hline
\multirow{2}{*}{{\bf Classifier}} &  \multicolumn{4}{c||}{{\bf Micro}} & \multicolumn{4}{c|}{{\bf Macro}}\\
\cline{2-9}
{}   & {\bf Precision}   & {\bf Recall} & {\bf F1} & {\bf ROC-AUC}    & {\bf Precision}   & {\bf Recall} & {\bf F1} & {\bf AUC} \\
\hline\hline
\botometer~ (RF) & 0.638 & 0.638 & 0.638 & 0.853 & 0.497 & 0.596 & 0.540 & 0.780\\
\spambot & 0.565 & 0.565 & 0.565 & 0.658 & 0.321 & 0.308 & 0.287 & 0.576\\
\fakeacc\ (LR) & 0.568 & 0.568 & 0.568 & 0.809 & 0.358 & 0.383 & 0.347 & 0.727\\
\ndsync\ (LR) & 0.573 & 0.573 & 0.573 & 0.764 & 0.383 & 0.293 & 0.269 & 0.545\\
\hline
Decision Tree & {\bf 0.754} & {\bf 0.754} & {\bf 0.754}  & 0.706 &  0.636 & 0.596 & 0.587  & 0.712\\
K-NN &  0.693 & 0.693 & 0.693 & 0.862 & 0.666 & 0.519 & 0.544 & 0.779\\ Logistic Regression &  {\bf 0.754} & {\bf 0.754} & {\bf 0.754}  & {\bf 0.924} & {\bf 0.731} & {\bf 0.641} & {\bf 0.671}  & {\bf 0.909}\\
Naive Bayes &  0.518 & 0.518 & 0.518 & 0.712 & 0.129 & 0.250 & 0.170 & 0.724 \\
SVM   &  0.746& 0.746 & 0.746  & 0.918 & 0.592 & 0.558 & 0.557 & 0.896 \\
Random Forest & 0.668 & 0.668 & 0.668 & 0.863 & 0.658 & 0.637 & 0.639 & 0.833 \\
Bagging   &  0.683 & 0.683 & 0.683 & 0.904 & 0.689 & 0.707 & 0.691 & 0.884 \\
Boosting   & 0.693 & 0.693 & 0.693  & 0.892 &  0.666 & 0.519 & 0.544 & 0.869\\\hline

\end{tabular}
\end{table*}

\begin{table*}
\centering
\caption{Performance of different competing methods for binary classification.}\label{table:binaryresult}
\begin{tabular}{|l||cccc||cccc|}
\hline
\multirow{2}{*}{{\bf Classifier}} &  \multicolumn{4}{c||}{{\bf Micro}} & \multicolumn{4}{c|}{{\bf Macro}}\\
\cline{2-9}
{}   & {\bf Precision}   & {\bf Recall} & {\bf F1} & {\bf ROC-AUC}    & {\bf Precision}   & {\bf Recall} & {\bf F1} & {\bf AUC} \\
\hline\hline
\botometer~ (DT) & 0.770 & 0.770 & 0.770 & 0.760 & 0.760 & 0.760 & 0.760 & 0.760\\
\spambot & 0.680 & 0.680 & 0.680 & 0.660 & 0.727 & 0.665 & 0.646 & 0.660\\
\fakeacc\ (SVM) & 0.696 & 0.696 & 0.696 & 0.701 & 0.704 & 0.701 & 0.693 & 0.701\\
\ndsync & 0.595 & 0.595 & 0.595 & 0.573 & 0.576 & 0.573  & 0.573 & 0.573\\
\hline
Decision Tree  &  0.859 & 0.859 & 0.859 & 0.633 &  0.640 & 0.632 & 0.622 & 0.632\\
K-NN &  0.799 & 0.799 & 0.799 & 0.795 & 0.814 & 0.795 & 0.795 & 0.795\\
Logistic Regression &  0.859 & 0.859 & 0.859  & 0.795 & 0.860 & 0.858 & 0.858  & 0.795\\
Naive Bayes &  0.518 & 0.518 & 0.518 & 0.500 & 0.259 & 0.500 & 0.341 & 0.500 \\
SVM   &  {\bf 0.873} & {\bf 0.873} & {\bf 0.873}  & {\bf 0.873} & {\bf 0.874} & {\bf 0.873} & {\bf 0.873} & {\bf 0.873} \\
Random Forest   &  0.782 & 0.782 & 0.782 & 0.786 & 0.797 & 0.785 & 0.780 & 0.786 \\
Bagging   &  0.695 & 0.695 & 0.695 & 0.698 & 0.709 & 0.698 & 0.684 & 0.698 \\
Boosting   & 0.799 & 0.799 & 0.799  & 0.873 &  0.814 & 0.795 & 0.795  & 0.873 \\\hline

\end{tabular}
\vspace{-5mm}
\end{table*}

\noindent{\textbf{(iv) \underline{Likelihood Features (LF):}}} If Twitter users want to publish tweets in a credit-based blackmarket service, they need to retweet a lot of others' tweets to earn credits for their own tweets to be retweeted by other customers. On the contrary, when they write tweets on Twitter and do not submit these tweets to any blackmarket services, they are not expected to perform such credit-based retweeting activity. We use the following set of features to capture this phenomenon: ({\bf LF\textsubscript{1-7}}) \textit{tweeting likelihood per day for seven days} ({\em Monday-Sunday}), ({\bf LF\textsubscript{8-14}}) {\em retweeting likelihood per day for seven days}, ({\bf LF\textsubscript{15-21}}) {\em regularity of tweeting activity per day for seven days}, ({\bf LF\textsubscript{22-28}}) {\em regularity of retweeting activity per day for seven days}, ({\bf LF\textsubscript{29}}) {\em tweet steadiness}, ({\bf LF\textsubscript{30}}) {\em retweet steadiness}, ({\bf LF\textsubscript{31-37}}) {\em maximum tweet likelihood per day for seven days}, and ({\bf LF\textsubscript{38-44}}) {\em maximum retweet likelihood per day for seven days}.  {\bf LF\textsubscript{1-7}} ({\em resp.} {\bf LF\textsubscript{8-14}}) is calculated by taking the ratio of the tweets ({\em resp.} retweets) of a user per day to the total number of tweets ({\em resp.} retweets) the user posted in a week. 
Regularity of tweeting activity per day ({\bf LF\textsubscript{15-21}}) is calculated by   
$-\sum_{\substack{i=1}}^{24} p(x_i)\log p(x_i)$, where $p(x_i)$ is the fraction of tweets posted by the user at $i^{th}$ hour of that day. We follow the same method to measure {\bf LF\textsubscript{22-28}} by replacing tweets with retweets. 
Tweet ({\em resp.} retweet) steadiness is calculated by $1/\sigma_t$ ({\em resp.} $1/\sigma_{rt}$) where $\sigma_t$ ({\em resp.} $\sigma_{rt}$) is the standard deviation of time difference between consecutive user-generated tweets ({\em resp.} retweets). {\bf LF\textsubscript{31-37}} ({\em resp.} {\bf LF\textsubscript{38-44}}) is calculated by the ratio of per-day tweet ({\em resp.} retweet) count of a user to the maximum number of tweets or ({\em resp.} retweets) the user posted in a day of a week.
 \\

\noindent{\textbf{(v) \underline{Fluctuation Features (FF):}}} Customers of credit-based blackmarket services want to acquire as many credits as possible in order to gain more retweets to their own tweets. We use the following features to validate this phenomenon: ({\bf FF\textsubscript{1}}) \textit{standard deviation of retweet counts for all user-generated tweets}, ({\bf FF\textsubscript{2}})  {\em mean of log-time difference between consecutive retweets}, ({\bf FF\textsubscript{3}}) {\em standard deviation of log-time difference between consecutive retweets}. 

Note that unlike \cite{cao2012aiding}, we did not use any graph-related features for two reasons: (i) crawling the complete neighborhood structure of users requires huge computational resources and sometimes produces incomplete information due to several constraints such as restriction of API and user profiles, (ii) given an unknown user account, collecting its neighborhood structure is time-consuming, which in turn may affect scalability of the real-time system we intended to build (Section \ref{sec:system}). 
However, we do not claim that graph-based features will not enhance the classification performance.  

\subsection{Classification Models}\label{sec:classifier}
We consider six state-of-the-art stand-alone supervised classifiers -- Decision
Tree (DT), K-Nearest Neighbors (K-NN), Logistic Regression (LR), Naive Bayes (NB), Support Vector Machine (SVM). We also consider three ensemble classifiers: Random Forest (RF), Bagging (BG) and Boosting (BO).
We perform hyper-parameter optimization in order to find the parameters that generate the best results. For instance, we use CART with Gini gain criteria for DT; K-NN method with K = 5; multinomial logistic regression and SVM with linear kernel.

\section{Experimental Results}
In this section, we start by briefly describing four baselines, followed by the detailed experimental results. 

\subsection{Baseline Methods}
Since there is no prior work on collusive retweeter detection, we choose four state-of-the-art methods as baselines, which are close to the problem we want to solve here. \\
{\bf Baseline I: \botometer:}
We use the social bot detection method proposed by \cite{davis2016botornot} and measure bot score for each retweeter. This score is further used as a feature to run different supervised classifiers mentioned in Section \ref{sec:classifier}.

\noindent\textbf{Baseline II: \spambot:}
We use the method proposed by \cite{wang2010detecting}  as our second baseline. It leverages a set of content-based and network-based features to detect spam bots from a collection of Twitter users. It assumes that {\em a genuine user is less likely to post duplicate tweets as compared to a bot}. 
To compare our method with this baseline, we use their proposed set of features and their suggested classifier (Naive Bayes) to classify retweeters as `Genuine', `Bots', `Promotional' and `Normal'.

\noindent\textbf{Baseline III: \fakeacc:} \cite{elazab2016fake} proposed a fake account detection model in Twitter based on
minimum weighted feature set. It first measures gain ratio of $22$ features and considers those features whose ratio crosses a certain threshold. Several classifiers are used, among which LR and SVM turned out to be the best for multi-class and binary classification respectively.    

\noindent\textbf{Baseline IV: \ndsync:} The closest baseline  of our method is \ndsync~ \cite{Neil:ndsync} which identifies multiple synchronous patterns across spam retweet threads. It extracts a set of features related to retweet threads (such as number of retweets, lifespan, response time etc.), projects retweet threads into a multi-dimensional feature space, and segments the feature space. It then estimates the suspiciousness score of each thread, and combines these scores for all the retweet threads associated with a user to obtain a user-level score. \ndsync\ is an unsupervised method to classify retweeters as `genuine' and `fake'. We use this method for binary classification. However, for multi-class classification, we utilize the suspicious user score returned by \ndsync~ as a feature for supervised classifiers.

\subsection{Evaluation Setup} We combine the customers collected from blackmarket services (see Table \ref{table:data_statistics}), yielding 743 customers (comprising of 86 bots,  275 promotional and 382 normal customers). We also sample 743 users randomly from a set of 1000 genuine users we collected, which makes the two classes balanced.
The accuracy of each competing method is measured using the following metrics: Precision, Recall, F1-score, and Area under the ROC curve (AUC). Since multi-class classification in involved, we report all these metrics in both Micro and Macro  settings (e.g., Micro-Precision, Macro-Precision etc.). We report the accuracy after averaging the result of 10-fold cross validation.

\begin{figure*}[t]
      \centering
      \includegraphics[width=2\columnwidth]{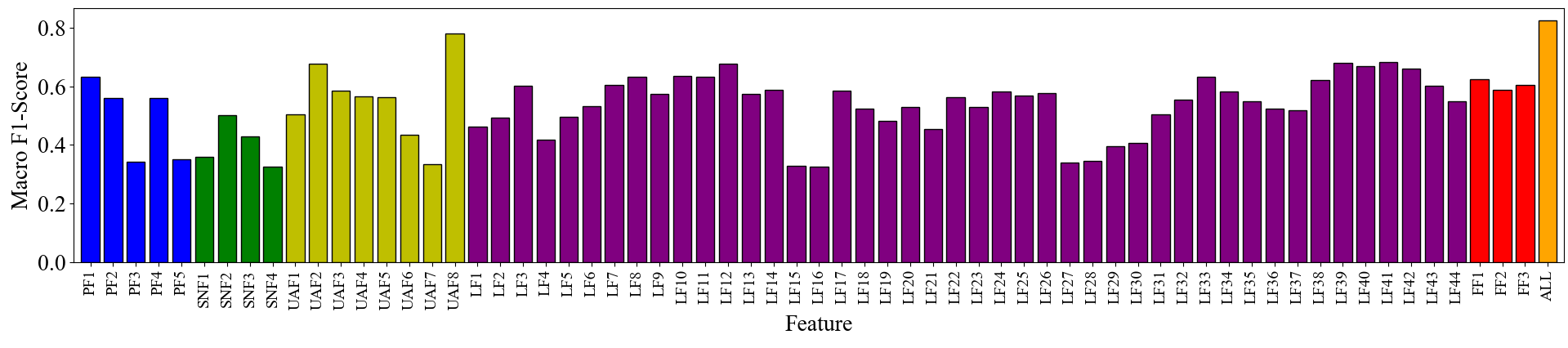}
      \caption{Feature importance (accuracy of SVM for binary classification considering each feature in isolation). We also plot the accuracy with all features at the right-most bar of the figure (labeled as `All'). }
      \label{fig:featureimportance}
      \vspace{-5mm}
\end{figure*}

\subsection{Results of Multi-class Classification} \label{sec:res_multi}
We design the first experimental setup as a four-class classification problem (genuine, bot, promotional, normal). 
 Table \ref{table:multiclassresult} shows the results for the four-class classification. We obtain the best result of \botometer~with Random Forest. Among four baselines, \botometer\ performs the best across all evaluation metrics. However, with our feature set,  Logistic Regression achieves the maximum accuracy -- it achieves 18\% and 8.3\% higher Micro-F1 and Micro AUC respectively, and 24.25\% and 15.38\% higher Macro-F1 and Macro AUC respectively w.r.t to the best baseline. The class-wise F1-score of Logistic Regression is as follows: 0.89 (genuine), 0.80 (bot), 0.58 (promotional), and 0.68 (normal).

\subsection{Results of Binary Classification} \label{sec:res_bin} One may only be interested to identify whether a retweeter is a customer or a genuine users, instead of a fine-grained classification of customers. Therefore, we conduct another set of experiments by combining all types of customers into a single class, and consider the problem as a binary classification problem. In Table \ref{table:binaryresult}, we notice that \botometer~with Decision Tree turns out to be the best baseline. However, this time SVM turns out to be the best model with a score of $0.873$ for all metrics, which is followed by Logistic Regression. SVM beats the best baseline by 11.3\% and 11.3\% in terms of  Micro-F1 and Micro AUC respectively, and by 10.3\% and 11.3\% in terms of Macro-F1 and Macro AUC respectively. The class-wise F1-score of Logistic Regression is as follows: 0.87 (genuine), 0.88 (customer).

\subsection{Feature Importance}
Figure \ref{fig:featureimportance} shows the importance of 64 features. Here we take each feature in isolation and run the best binary-classification model (SVM)\footnote{The pattern of feature importance is same for the multi-class classification with Logistic Regression.}. The most important feature seems to be Bot-score (UAF\textsubscript{8}) with which SVM achieves a Macro F1-score of 0.75. It also corroborates with the significantly high accuracy of our first baseline model (\botometer). The second and third ranked features are -- maximum retweet likelihood in every Tuesday (LF\textsubscript{39}) and in every Friday (LF\textsubscript{42}) respectively. One possible reason why LF\textsubscript{39} and LF\textsubscript{42} have better discriminatory power is that these blackmarket services refresh their tweet database every 3-4 days (possibly every Tuesday and Friday). Thus, in order to keep their credit high, customers must keep on retweeting tweets every 3-4 days. However, as a whole, fluctuation features turn out to be the best (Macro F1=0.61), followed by user activity feature (Macro F1=0.554) and likelihood feature (Macro F1=0.55).   

\begin{figure}[t]
      \centering
      \fbox{\includegraphics[width=0.9\columnwidth]{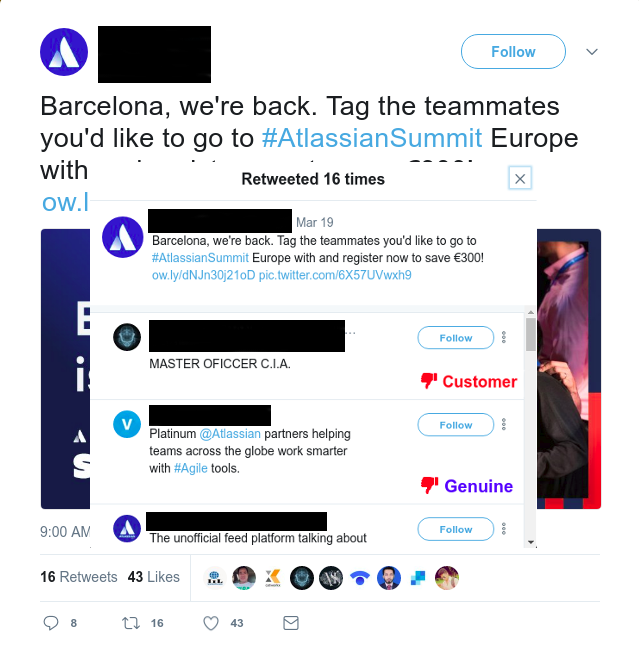}}
      \caption{Browser extension: \score.}
      \vspace{-5mm}
      \label{fig:browser}
\end{figure}

\section{Browser Extension: SCoRe}\label{sec:system}
In the previous section, we have showed that state-of-the-art supervised models are capable of producing a significant accuracy with our proposed feature set to segregate blackmarket customers from genuine users. In order to help users deal with such fake retweeting activities, we attempt to build an extension, named \score~for chrome browser. It allows users to spot the blackmarket customers, which in turn provides a better way to understand the importance of a tweet. 

Figure \ref{fig:browser} shows a snapshot of how \score~facilitates users. The input is a link of the tweet whose retweeters the user wants to analyze. It first extracts all its retweeters, calculates the features  for each retweeter, and then feeds it to the pre-trained SVM. Finally, the label (Genuine or Customer) of each retweeter is displayed. It also allows users to provide feedback on each label. There is a `thumbs down' symbol associated with each label, upon clicking of which the feedback will be forwarded to the back-end server. \score~has the capability to be trained incrementally with the feedback provided by the users. However, special care has been taken to make the learning process robust by ignoring spurious feedback. An attacker may want to pollute \score~ by injecting wrong feedback. \score~handles these spurious feedback by first checking the confidence of the current model on labels associated with the feedback, and ignoring them if the confidence is more than a pre-selected threshold (currently set as 0.75). This makes \score~more robust under adversarial attacks.

\noindent{\bf User study:}
To analyse the performance of \score, we conducted a user study with the help of 25 volunteers. First, we randomly assigned 20 tweet URLs to each volunteer. Volunteers visited each URL with the extension activated in their browser, clicked on the retweeter list (each retweeter was labeled as `Genuine' or `Customer' by \score) and marked the wrong label by pressing `thumbs down' button. \score~took 10 sec to label each retweeter. We recorded the responses of the volunteers and measured the average accuracy. The results shows that \score~ is highly accurate, achieving an average accuracy) of $85\%$ (with standard deviation of $0.02$).

\begin{figure}[!t]
      \centering
      \includegraphics[width=\columnwidth]{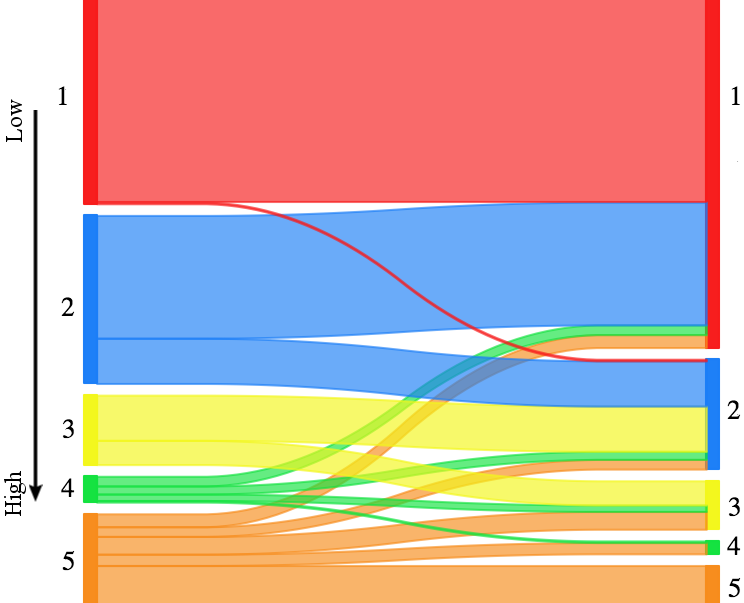}
      \caption{{\bf Alluvial diagram representing the flows of tweets between the bins (based on retweet count) before and after filtering customers.} The colored blocks correspond to different bins (the range of retweet count is divided equally into five bins). Left (right) blocks correspond to bins based on retweet rank before (after) the filtering. The size of the block indicates the number of tweets in that bin, and the shaded waves joining the regions represent flow of tweets between the bins, such that the width of the flow corresponds to the fraction of tweets.}
      \label{fig:alluvial}
      \vspace{-5mm}
\end{figure}

\section{Implication of collusive Retweeer Detection}
Once we detect collusive retweeters, its immediate implication would be to re-rank the tweets based on modified retweet count (after filtering blackmarket customers). Figure \ref{fig:alluvial} shows an alluvial diagram, indicating how tweets (500 tweets collected from the customers' timelines) change their ranking after such filtering.  We show that top- and middle-ranked tweets are mostly affected by this filtering.  We believe that this analysis opens the scope for a serious re-investigation of the existing metrics for ranking tweets/users.

\section{Conclusion}
In this paper, we studied the problem of understanding and detecting blackmarket-based collusive retweeters. The major contributions of this work are fourfold: {\bf  Dataset:} We collected a dataset of blackmarket collusive retweeters and annotated them. This, to our knowledge, is the first dataset of such kind.  {\bf Characterization:} We propose 64 novel features to characterize customers and differentiate them from normal users. {\bf Classification:} State-of-the-art supervised methods performed significantly well to classify customers and genuine retweeters.  {\bf System design:} We developed \score, a chrome extension that spots blackmarket retweeters in real-time.

\section*{Acknowledgement}
The work was partially supported by Flipkart, India,  Ramanujan Faculty Fellowship, and the Infosys Center for AI.

\bibliographystyle{abbrv}
\bibliography{ref}

\end{document}